\documentclass[usedcolumn,usenatbib,usegraphicx]{mn2e}

\title[Long-term evolution of accretion discs in Be/X-ray binaries]
      {Long-term evolution of accretion discs in Be/X-ray binaries}

\author[Kimitake~Hayasaki and Atsuo~T.~Okazaki]
  {Kimitake~Hayasaki$^{1,2}$\thanks{E-mail: hayasaki@topology.coe.hokudai.ac.jp}
   and Atsuo~T.~Okazaki$^3$ \\
  $^1$Department of Applied Physics, Graduate School of Engineering,
        Hokkaido University, Kitaku N13W8, Sapporo 060-8628, Japan.\\
  $^2$Centre for Astrophysics and Supercomputing, Swinburne University of Technology,
        Hawthorn Victoria 3122 Australia.\\
  $^3$Faculty of Engineering, Hokkai-Gakuen University, Toyohira-ku,
      Sapporo 062-8605, Japan.}
\date{}

\pagerange{\pageref{firstpage}--\pageref{lastpage}} \pubyear{2004}

\def\LaTeX{L\kern-.36em\raise.3ex\hbox{a}\kern-.15em
    T\kern-.1667em\lower.7ex\hbox{E}\kern-.125emX}

\begin{document}

\label{firstpage}

\maketitle

\begin{abstract}

We numerically study the long-term evolution of the accretion disc
around the neutron star in a coplanar Be/X-ray binary with a short
period and a moderate eccentricity. From three dimensional Smoothed
Particle Hydrodynamics simulations, we find that the disc evolves through
three distinct phases, each characterized 
by different mass accretion patterns. 
In the first "developing phase", 
the disc is formed and develops towards
a nearly Keplerian disc. It has a relatively large, double-peaked
mass-accretion rate with the higher peak by the direct accretion at
periastron, which is followed by the lower peak by the accretion
induced by a one-armed spiral wave.
In the second "transition phase", the disc is approximately Keplerian and grows with
time. The mass-accretion rate increases as the disc grows.
In the second phase, there is a transition in the mass accretion rate from 
a double peaked to a single peaked pattern.
In the final quasi-steady state, the
mass-accretion rate is on average balanced with the mass-transfer rate
from the Be disc and exhibits a regular orbital modulation. In the
quasi-steady state, the mass-accretion rate has a single peak by the
wave-induced accretion as in a later stage of the transition phase.
The orbital modulation of X-ray maxima could provide not only a
circumstantial evidence for the persistent disc but also an
observational diagnosis of the disc evolutionary state.

\end{abstract}


\section{Introduction}
\label{sec:intro}

The Be/X-ray binaries comprise the largest subclass of high-mass
X-ray binaries (HMXBs). About two-thirds of the identified systems fall into
this category. These systems consist of, generally, a neutron star and
a Be star with a cool ($\sim 10^{4}K$) equatorial disc, which is
geometrically thin and nearly Keplerian. 
These systems are distributed over a wide range of orbital periods
$(10\,{\rmn{d}} \la P_{\rm{orb}} \la 300\,{\rmn{d}})$ and eccentricities
$(e \la 0.9)$.

Most of the Be/X-ray binaries show only transient activity in the
X-ray emission and are termed Be/X-ray transients. Be/X-ray transients
show periodical (Type I) outbursts, which are separated by the orbital
period and have the luminosity of
$L_{\rmn{X}}=10^{36-37}\rmn{erg\,s}^{-1}$, and giant (Type II)
outbursts of $L_{\rmn{X}} \ga 10^{38} \rmn{erg\,s}^{-1}$ with no
orbital modulation. 
These X-ray outbursts are basically considered to result from the 
accretion onto the neutron star from the Be disc.

Recently, \citet{haya} (hereafter, Paper~I) studied the accretion flow around 
the neutron star in Be/X-ray binaries, performing three-dimensional (3D)
smoothed particle hydrodynamics (SPH) simulations (\citealt{be1}; \citealt*{ba}),
 in the framework of the interaction between the
Be disc and the neutron star (\citealt{oka2}).
They found that the time-dependent 
accretion disc is formed around the neutron star. 
\citet{haya2} (hereafter, Paper~II) further showed that the disc has a
one-armed spiral structure induced by a phase-dependent mass transfer from the Be disc.
The one-armed mode is excited 
and then damped every orbit, which make an enhancement in the X-ray luminosity.
These are, however, the results from simulations
run over a period shorter than the viscous time-scale of the disc,
and it was not known how the accretion disc would evolve over a period
longer than the viscous time-scale.

In this paper, we study the long-term evolution of the accretion disc around 
the neutron star in Be/X-ray binaries, using a 3D SPH code.
The simulations are performed using the basically the same method as in paper~I.
We describe our numerical results in Sections 2 and 3, and
discuss the disc evolution and its observational implications in Section 4.
Section~5 summarizes our conclusions.


\begin{table*}
\caption{Summary of model simulations.
  The initial number of
  particles is 301 and the mean mass injection rate is
  $\sim2.5\times10^{-11}\rho_{-11}M_{\odot}\rmn{yr}^{-1}$ in all
  simulations, where $\rho_{-11}$ is the highest local 
  density in the Be-star disc normalized by $10^{-11}{\rm g,cm}^{3}$. 
  The first column represents the model. The second
  column is the polytropic exponent $\Gamma$ and the third column is
  the number of SPH particles at the end of the run.
  In the fourth column,
  the inner radius of the simulation region is described.
  The last column is the mean accretion rate
  for the last one orbital period.
}
\label{tbl:models}
\begin{tabular}{@{}lcccc}
\hline
Model  & Polytropic exponent & $N_{\rmn{SPH}}$
       & Inner boundary
       & $\dot{M}_{\rmn{acc}}$ \\
       & $\Gamma$        & (final)
       & $r_{\rmn{in}}/a$
       & $(\rho_{-11}M_{\odot}\rmn{yr}^{-1})$ \\
\hline
1      & $1.2$ & $66676$
       & $3.0\times10^{-3}$
       & $1.7\times10^{-11}$\\
2      & $5/3\hspace{1mm}(\rmn{adiabatic})$ & $22122$
       & $3.0\times10^{-3}$
       & $2.5\times10^{-11}$\\
3      & $1\hspace{1mm}(\rmn{isothermal})$ & $49950$
       & $1.0\times10^{-2}$
       & $1.5\times10^{-11}$\\
\hline
\end{tabular}
\end{table*}


\section{Disc structure}
\label{sec:dstruc}

We have carried out simulations by using the same computer code as in
papers~I and II. In our code, the neutron star is modeled by a sink
particle with a fixed accretion radius $r_{\rmn{in}}$, while the
accretion disc is modeled by an ensemble of non-selfgravitating gas
particles. As in papers~I and II, our simulations are for a coplanar
system with a short period ($P_{\rmn{orb}}=24.3\,\rmn{d}$) and a
moderate eccentricity $(e=0.34)$, the parameters of which are for 4U\,0115+63, one
of the best studied Be/X-ray binaries. The phase-dependent,
mass-transfer rate from the Be disc is emulated by injecting gas
particles, based on the result from a high resolution SPH simulation
for the same binary configuration \citep{oka2}. The SPH
particles are assumed to have the initial temperature of half the
effective temperature of the Be star and follow a polytropic equation
of state specified by the exponent $\Gamma$.

We consider the case $\Gamma=1.2$ (model~1) as well as the adiabatic
case $\Gamma=5/3$ (model~2) and the isothermal case $\Gamma=1$
(model~3) as two extreme cases. Models~1-3 in this paper correspond
to models~1, 3 and 4 in paper~I, respectively.  The injection rate of
the SPH particles in each model is three times as low as in
corresponding simulation in paper~I in order to make the computing
time bearable. In models~1 and 2, we adopt $r_{\rmn{in}} =
3.0\times10^{-3} a$, where $a$ is the semi-major axis of the
binary. In model~3, however, a larger accretion radius of
$r_{\rmn{in}}=1.0\times10^{-2} a$ is adopted, in order to reduce an
artificial effect of the inner simulation boundary.
All models have the Shakura-Sunyaev viscosity parameter
$\alpha_{\rm SS}=0.1$ and are run for $45 P_{\rm orb}$, 
which is twice, 10 times and 5 times as long as the
viscous time-scale at the disc outer radius in models~1, 2 and 3,
respectively (see Section 3.3). The number of neighbor particles is
always kept to about 50 particles throughout our simulations
Details of the models are given in Table~1.
Throughout this paper, the unit of time is $P_{\rmn{orb}}$, unless
noted otherwise.

We perform three dimensional (3D) simulations, despite that they
have less radial resolution than two dimensional (2D) simulations do
with the same particle number. 3D calculations have an advantage
even for a geometrically thin disc in a coplanar binary. They enable
us to evaluate more accurately the effect of the one-armed wave
excited by the impulsive mass transfer from the Be disc than 2D
calculations do. As we will see in later sections, the one-armed
wave plays an important role in the long-term evolution of the
accretion disc. 2D simulations would have artificially enhanced
wave amplitude because of the lack of vertical structure 
(e.g., \cite{og}).

In this section, we show the azimuthally averaged structures, 
the non-axisymmetric structures and the vertical structures 
of the fully developed discs during the last
one orbital period $44\le{t}\le45$.


\subsection{Azimuthally averaged disc structure}
\label{sec:aastruc}

\begin{figure*}
\resizebox{\hsize}{!}{
\includegraphics{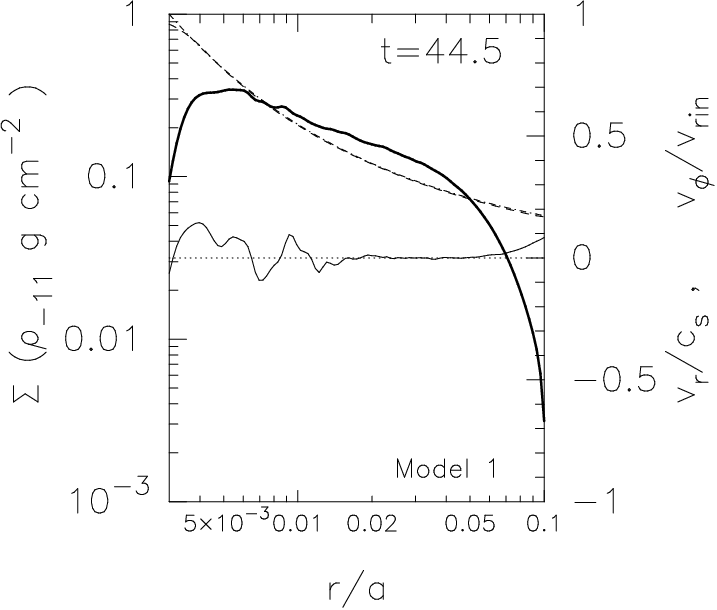}
\includegraphics{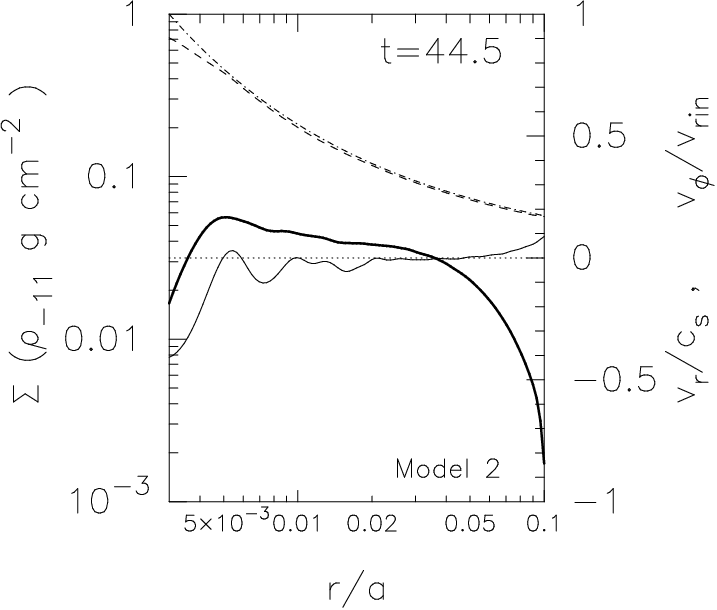}
\includegraphics{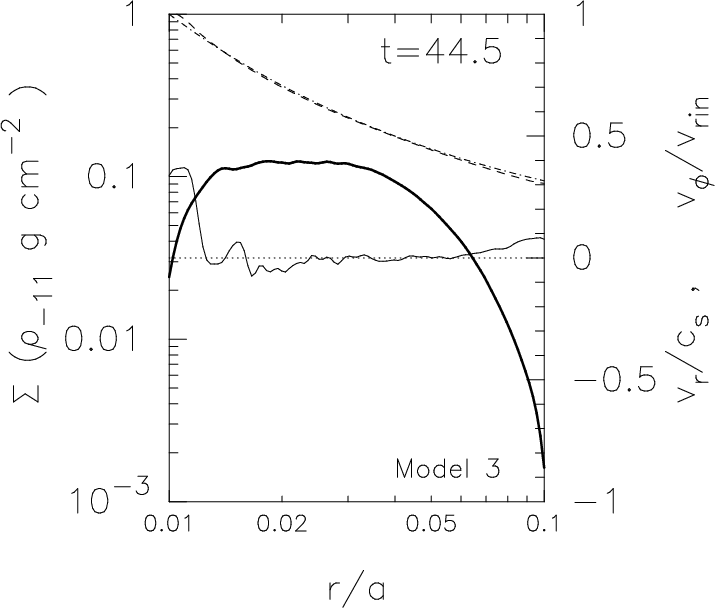}}
 \caption{Azimuthally averaged radial structures of the accretion disc at \rm{t}=44.5 \
in models~1,2 and 3. In each panel, the thick solid line, the dashed line and the thin solid line 
denote the surface density $\Sigma$ in units of $\rho_{-11}\rm{g\,cm^{2}}$, the azimuthal velocity
$v_{\phi}$ normalized by the Keplerian velocity at the inner boundary and the radial Mach number 
$v_{r}/c_{s}$, respectively. For comparison, the Keplerian velocity distribution is shown by 
the dash-dotted line. 
}
 \label{fig:discst}
\end{figure*}

Fig.~\ref{fig:discst} shows the radial structure of the accretion disc around the 
neutron star at apastron ($t=44.5$).
In each panel, the thick solid line, the dashed line and 
the thin solid line denote the surface density $\Sigma$, 
the azimuthal velocity $v_{\phi}$ normalized by the Keplarian 
velocity at the inner boundary and the radial Mach number $v_{r}/c_{s}$,
respectively.
In the figure, the velocity components are averaged vertically and azimuthally,
 while the density is integrated vertically and averaged azimuthally, where
$\rmn{\rho_{-11}}$ is highest local
density in the Be-star disc normalized by $\rmn{10^{-11}} \rmn{g}
\rmn{cm^{-3}}$, a typical value for Be stars.
It is noted from Fig.~\ref{fig:discst} that the disc is nearly Keplerian and 
the radial velocity component of the disc is highly subsonic in each model.
It is also noted that the surface density in model~2 (adiabatic case) is the lowest
because of the highest viscous stress due to the highest pressure.

\begin{figure*}
\resizebox{\hsize}{!}{
\includegraphics*[width=0.4cm]{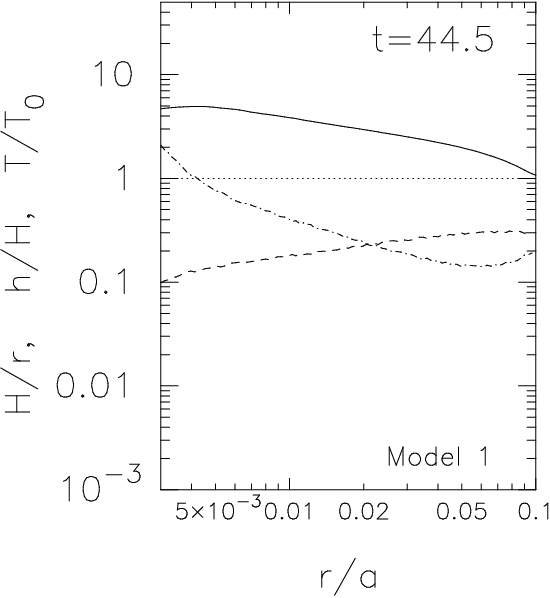} 
\includegraphics*[width=0.4cm]{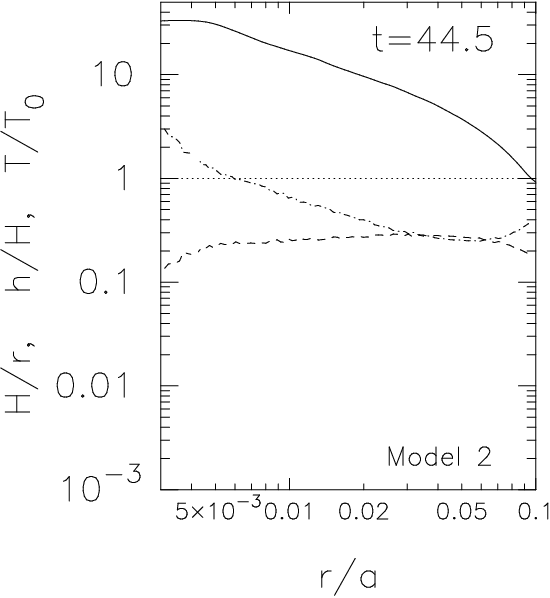} 
\includegraphics*[width=0.4cm]{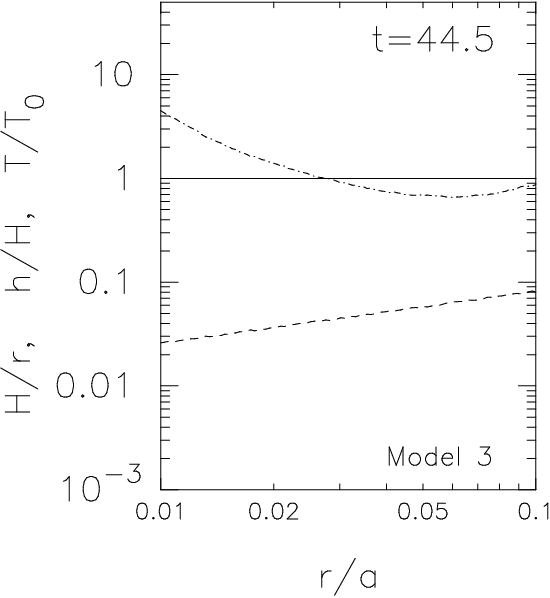}} 
 \caption{Radial distributions of some characteristic quantities at \rm{t}=44.5 \
in models~1,2 and 3. In each panel, the solid line, the dashed-dotted line 
and the dash line denote the disc temperature $T$ normalized by $T_{0}=1.3\times10^{4}\rm{K}$
 which is the initial temperature of injected particles, 
the ratio of the smoothing length to 
the disc thickness $h/H$ and 
the ratio of the disc thickness to the radius $H/r$, respectively.} 
 \label{fig:discst2}
\end{figure*}

Fig.~\ref{fig:discst2} denotes the radial structures of the disc temperature
, the ratio of the smoothing length to the disc thickness
and the ratio of the disc thickness to the radius at $t=44.5$ in models~1-3.
In each panel, the solid line, the dashed-dotted line and the dashed line 
denote the disc temperature $T$ normalized by $T_{0}=1.3\times10^{4}\rm{K}$, 
which is the initial temperature of injected particles, the ratio of 
the smoothing length to the disc thickness $h/H$ and 
the ratio of the disc thickness to the radius $H/r$, respectively.
Here the disc thickness $H$ is determined from the vertical
distribution of the SPH particles by comparison with the analytical
density distribution of a geometrically thin disc. The vertical disc
structure and the procedure to determine $H$ from the particle
distribution will be given in Section~\ref{sec:vertical}.
It is noted from the figure that the disc is 
fully resolved in the radial direction and is also resolved
in the vertical direction except in the innermost region of the disc.

In our simulations, the presence of the inner simulation boundary
causes an artificial decrease in the density near the boundary,
because the gas particles which pass through the boundary are removed
from the simulation. As a result, the density distribution has a break
near the inner boundary, as seen in Fig.~\ref{fig:discst}.
We have found that our simulations are reliable for $\ga 2 r_{\rmn{in}}$.


\begin{figure*}
\resizebox{\hsize}{!}{
\includegraphics*[height=0.395cm,width=0.4cm]{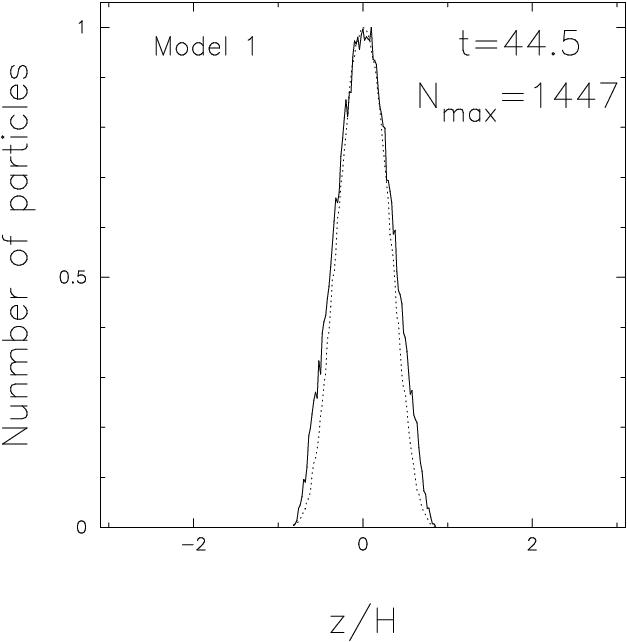}
\includegraphics*[height=0.395cm,width=0.4cm]{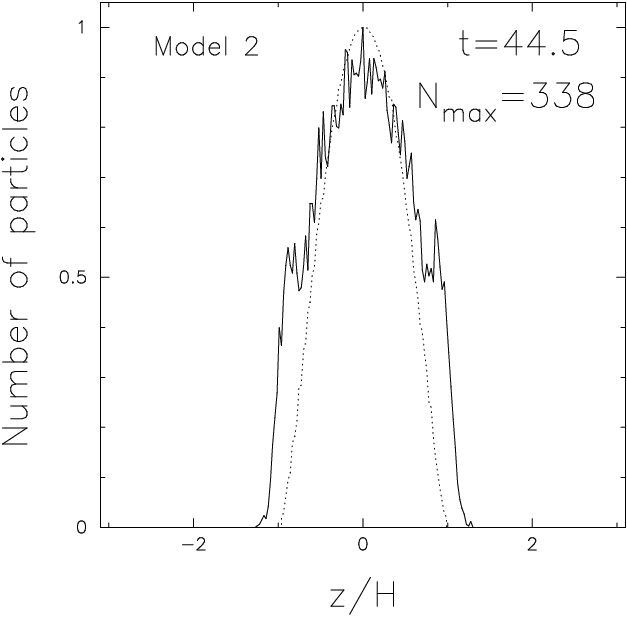}
\includegraphics*[height=0.395cm,width=0.4cm]{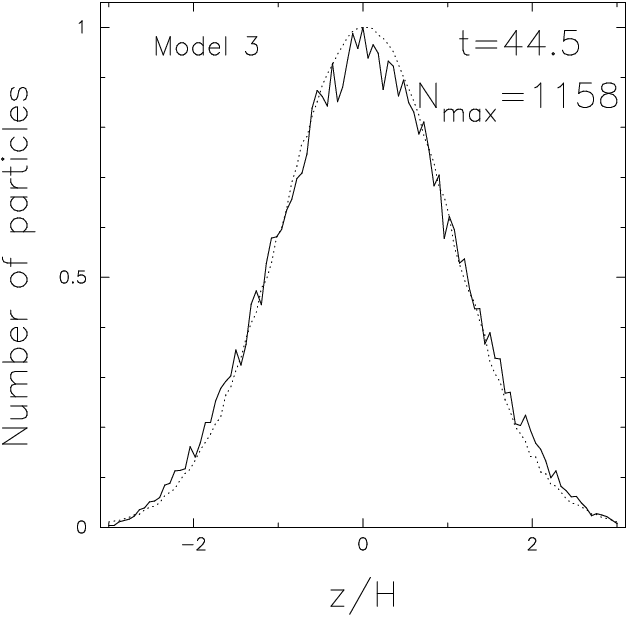}}
 \caption{
Particle distributions in the vertical direction
at \rm{t}=44.5 
in models~1-3. 
In each panel, the solid line
and the dotted line denote the number of SPH particles normalized 
by the number of particles $N_{\rm{max}}$ at the disc mid-plane and 
the corresponding analytical distribution, respectively.
}
 \label{fig:discst2a}
\end{figure*}


\subsection{Non-axisymmetric disc structure}
\label{sec:nonaxis}

It was found by paper~II that an accretion disc around
the neutron star in Be/X-ray binaries has a one-armed spiral
structure induced by a phase-dependent mass transfer
from the Be disc to the neutron star.
In this subsection, 
we show the non-axisymmetric structure of 
the accretion disc for models~1-3.

\begin{figure}
\resizebox{\hsize}{!}
{
\includegraphics*{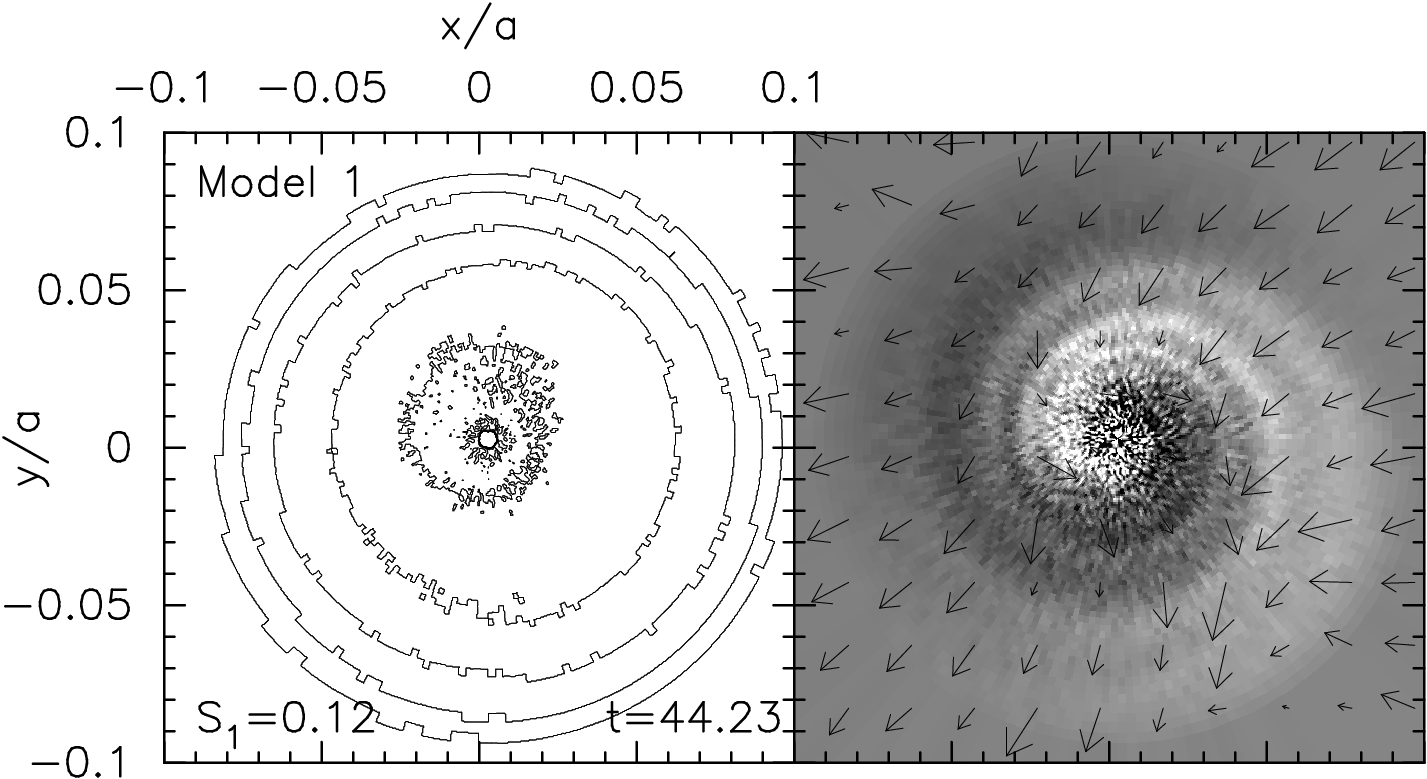}}\\
\resizebox{\hsize}{!}
{
\includegraphics*{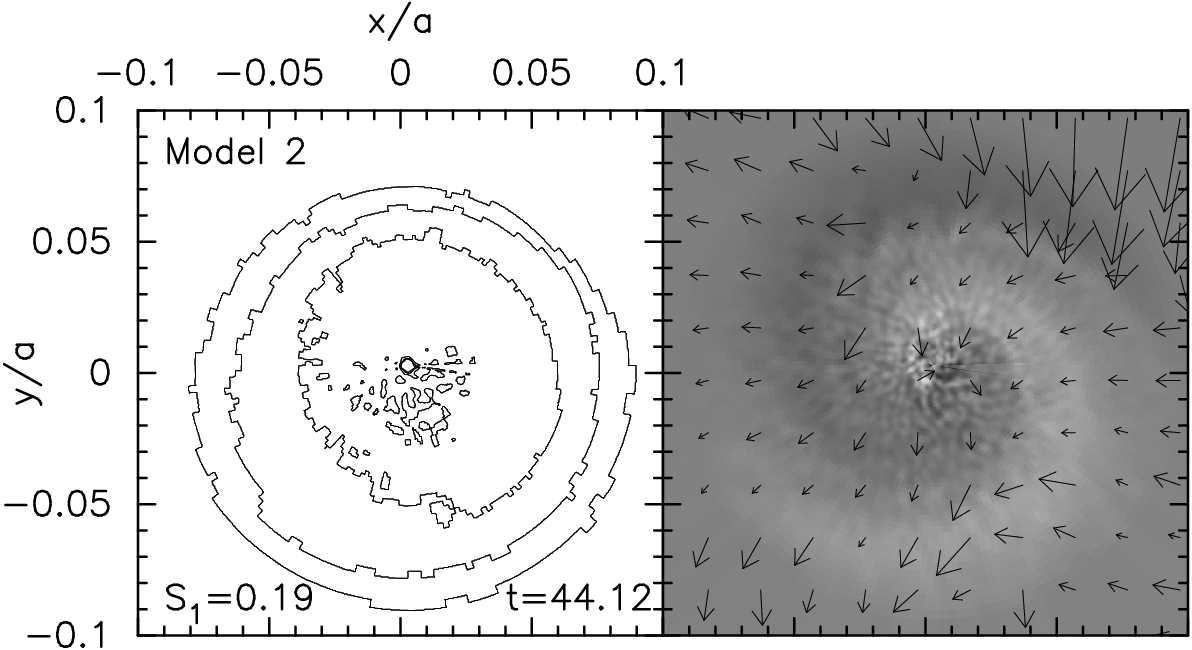}}\\
\resizebox{\hsize}{!}
{
\includegraphics*{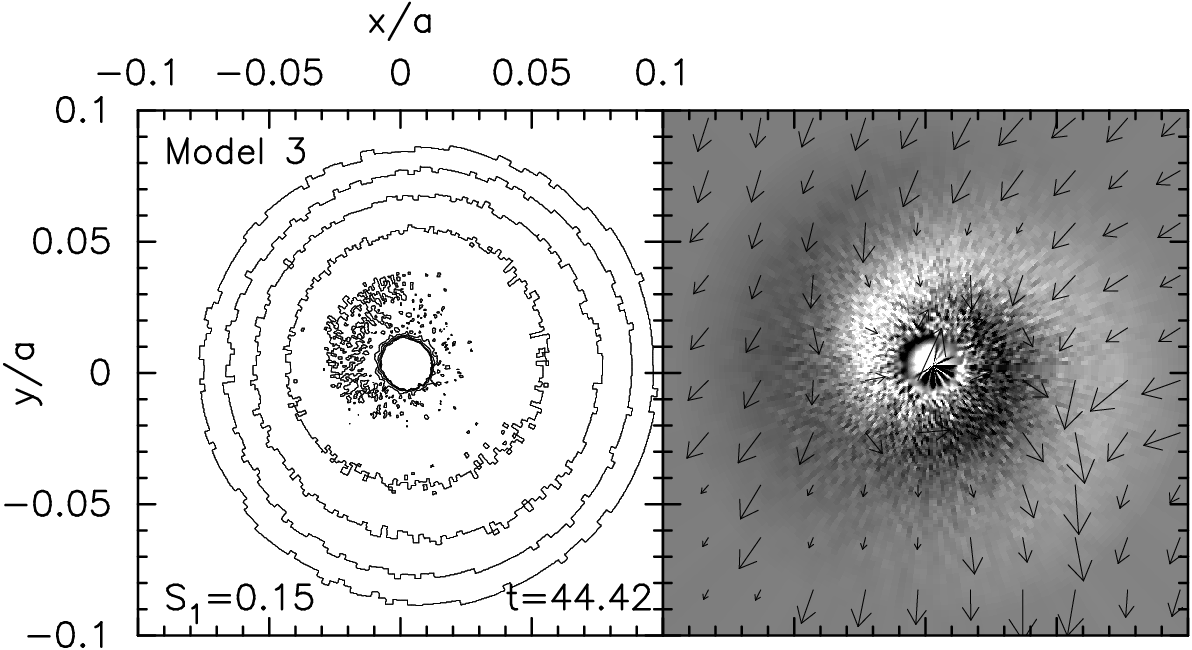}}
 \caption{
   Snapshots of the accretion disc for models~1-3.
   In each figure, the left panels show the surface density in a range of
   five orders of magnitude in the logarithmic scale,
   while the right panels show the non-axisymmetric components of
   the surface density (gray-scale plot)
   and the velocity field (arrows) in the linear scale.
   In the right panels, the region in gray (white) denotes the region
   with positive (negative) density enhancement.
   The periastron is in the $x$-direction and
   the disc rotates counterclockwise.
   Annotated in each left panel are the time
   in units of $P_{\rm orb}$ and the mode strength $S_{1}$.}
 \label{fig:discst3}
\end{figure}

Fig~\ref{fig:discst3} gives snapshots of the accretion disc 
around the neutron star at $t=44.23$ in model~1, at $t=44.12$ in model~2 
and at $t=44.43$ in model~3, 
respectively.
Each time $t$ is chosen such that the strength of the one-armed 
mode takes a maximum over the period of $44\le{t}\le45$.
The left panels show the contour maps of the surface density, 
whereas the non-axisymmetric components of the surface density 
and the velocity field are shown in the right panels. 
Annotated in each left panel are the time in units of $P_{\rm{orb}}$ 
and the amplitude of the one-armed wave $S_{1}$, details of which will be described later.
It is noted from the figure that the disc has a one-armed spiral structure 
regardless of the simulation parameters, even after the disc fully evolves.
The figure also suggests that the strength of one-armed mode 
correlates inversely with the surface density.


\subsection{Vertical disc structure}
\label{sec:vertical}
 
In this subsection, we compare the vertical distribution of SPH 
particles with the analytical density distribution of a 
geometrically thin disc.
The vertical structure of such a disk is characterized by the disc 
scale-height ($¥Gamma=1$) or thickness ($¥Gamma ¥ne 1$).
In an isothermal disc, the vertical scale-height is analytically 
given by
\begin{equation}
H_{\rm{a}}=c_{s}/\Omega_{K},
\label{eq:discst2a2}
\end{equation}
whereas the disc thickness with the polytropic equation of state
is defined by
\begin{equation}
H_{\rm{a}}=\rho_{0}^{(\gamma-1)/2}((\gamma-1)/2\gamma K_{\rm p})^{-1/2}\Omega_{K}^{-1},
\label{eq:discst2a3}
\end{equation}
where $c_{s}$ is the sound speed, $\Omega_{K}$ is the Keplerian frequency, 
$\rho_{0}$ is the central density of the disc and
$K_{\rm p}$ is the polytropic constant.

Fig.~\ref{fig:discst2a} shows the SPH particle distributions
in the vertical direction in models~1-3.
In each panel, the solid line and the dashed line denote the
 number of $\rm{SPH}$ particles normalized 
by the number of particles $N_{\rm max}$ at the disc mid-plane
and the corresponding analytical distributions with $H_{\rm{a}}$ 
given by equations~(\ref{eq:discst2a2}) or (\ref{eq:discst2a3}).

It is noted from the figure that
the distributions of $\rm{SPH}$ particles are 
slightly broader than those of analytical distributions 
except for model~3 (isothermal disc).
The ratio of the disc thickness obtained by 
the $\rm{SPH}$ particle distribution
to the disc scale-hight or thickness by analytical ones $H/H_{\rm{a}}$ is
1.12 in model~1, 1.32 in model~2 and 1.03 in model~3.
The reason why the isothermal model has a better fit than models 
with $¥Gamma ¥ne 1$ do is that the scale-height of the isothermal
disc depends only on the temperature, while models 
with $¥Gamma ¥ne 1$ depend on the density, which has a numerical error.
In any case, discs in our simulations are vertically resolved 
except in the innermost region.


\section{Disc evolution}
\label{sec:devolve}

By analysing the
multi-wavelength long term monitoring observations of 4~U0115+63,
 \cite{ne} found that
the Be star
shows a quasi-cyclic activity ($\simeq3-5\rm{yr}$) due to the loss and 
reformation of its circumsteller disc.
When the Be disc is lost,
the mass-supply to the neutron star also has to halt.
Since we have little knowledge on the mechanism that triggers the
Be-disc loss episode, we have restricted our simulations for
$t \la3\,{\rm yr}\sim 45\,P_{\rm{orb}}$.

In the first two subsections below, 
we will describe the evolution of the mass-accretion rate and
the strength of modes for models~1-3. 
Then, we will show the evolution of the mass, radius and 
viscous time-scale of the accretion disc for these models.


\subsection{Mass accretion rate}
\label{sec:accretion}

\begin{figure*}
\resizebox{\hsize}{!}{\includegraphics*{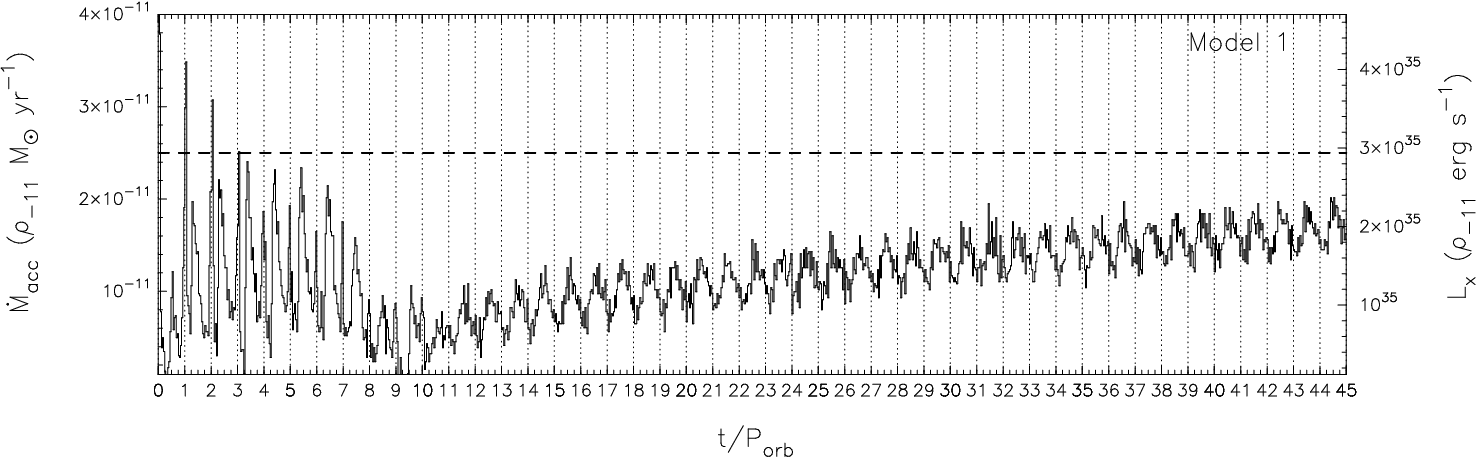}}
\resizebox{\hsize}{!}{\includegraphics*{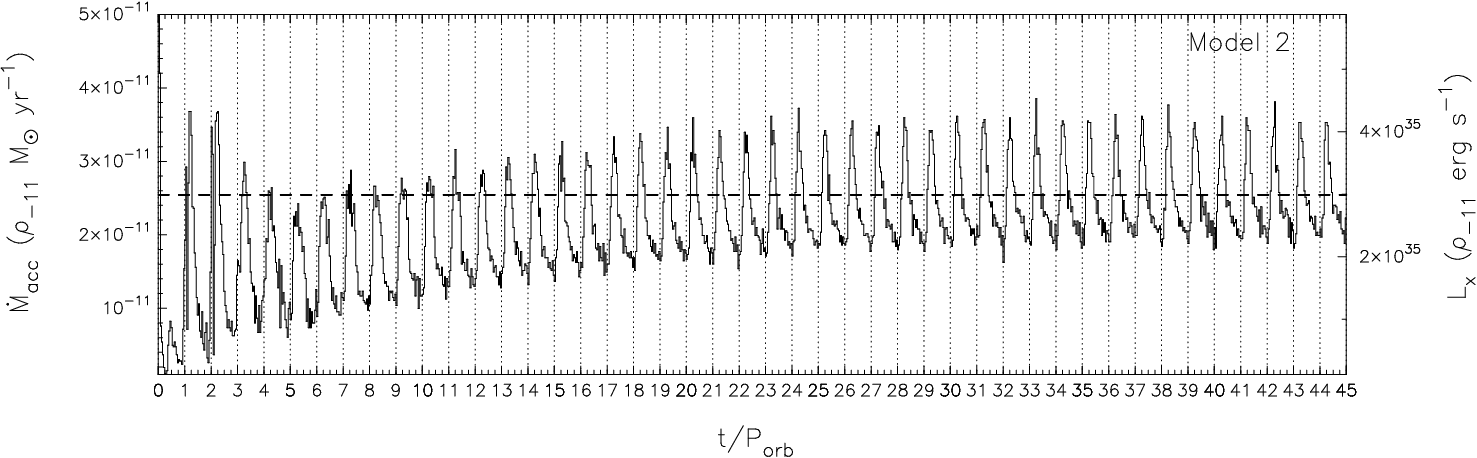}}
\resizebox{\hsize}{!}{\includegraphics*{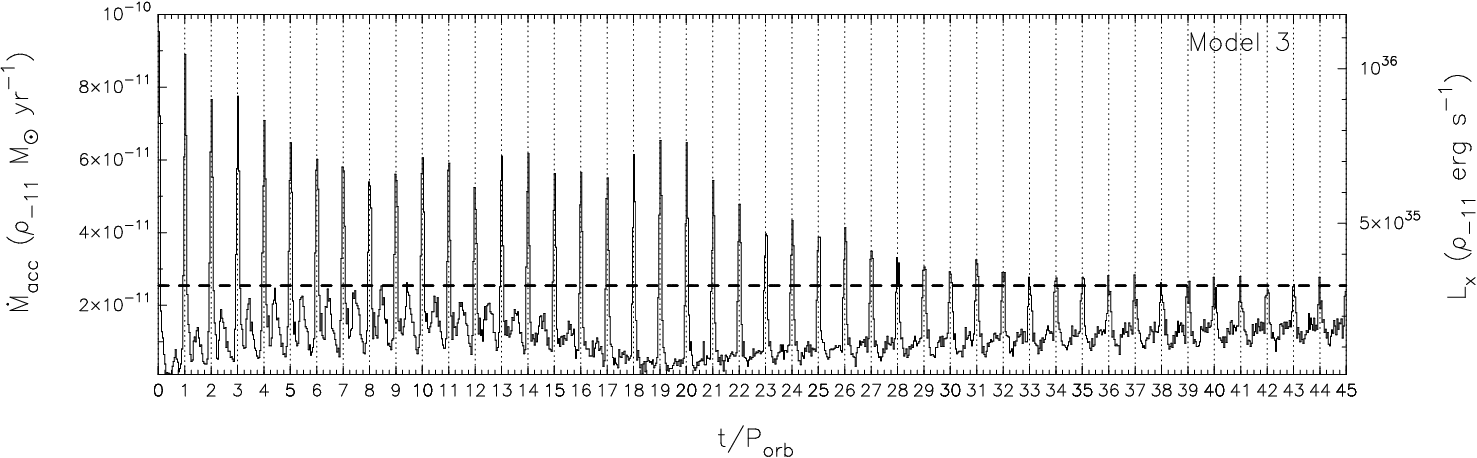}}
 \caption{
   Evolution of the mass-accretion rate $\dot{M}_{\rmn{acc}}$
   for models~1-3 with different polytropic exponents. 
   In each panel, the solid line and the dashed horizontal line denote the
   mass-accretion rate and the averaged mass-transfer rate from the Be disc, respectively. The
   mass-accretion rate is measured in units of $\rho_{-11}
   M_{\odot}\,\rmn{yr}^{-1}$. 
   The right axis shows
   the X-ray luminosity corresponding to the mass-accretion rate with the
   X-ray emission efficiency $\eta=1$, where $\eta$ is defined by
   $L_{X}=\eta M_{\rmn{X}}\dot{M}_{\rmn{acc}}/R_{\rmn{X}}$.
          }
 \label{fig:mdots}
\end{figure*}

Fig.~\ref{fig:mdots} shows the evolution of the mass-accretion
rate and the corresponding X-ray luminosity in models~1-3 for $0 \le t
\le 45$. In each panel, the solid line and the dashed line 
denote the mass-accretion rate $\dot{M}_{\rm{acc}}$ 
and the mean mass-transfer rate $\dot{M}_{\rm{T}}$
, respectively, where $\dot{M}_{\rm{T}}$ is 
$\simeq2.5\times10^{-11}\rho_{-11}M_{\odot}yr^{-1}$.
$\dot{M}_{\rm{acc}}$ in models~1, 2 and 3 are shown in 
the top panel, the middle panel and the bottom panel, respectively.

In model~1 (top panel), 
the mass accretion rate has double peaks per orbit 
at an initial developing phase $0\le{t}\la10$. 
While the first peak is due to the direct accretion of particles 
with the low specific angular momentum, 
the second one is mainly caused by an inward propagation of $m=1$ mode. 
After the disc is fully developed ($t\ga10$), the mass accretion rate 
has a single peak per orbit only due to the wave induced accretion.
In the long-term, the mass accretion rate gradually increases with time 
after $t=10$

In model~2 (middle panel), 
the mass accretion rate basically
shows the same behaviour as that of model~1 
in the short term, 
except that 
its peak value is higher than
the mean mass-transfer rate after the disc is developed.
In the long term, the mass accretion rate gradually increases with time 
and saturates for $t\ga20$ with an orbital modulation.
This is because the mass of gas transferred from the Be disc over one orbital period
is balanced with that of the accreting gas on to the neutron star
for $t\ga20$. For example, as shown in Table~\ref{tbl:models}, 
the mean mass-accretion rate for the last one orbital period is
$2.5\times10^{-11}\rho_{-11}M_{\odot}yr^{-1}$, which  
approximately equals to the mean mass-transfer rate from the Be disc.
Thus, we noted that the disc gets to the quasi-steady state for $t\ga20$ in model~2.

In model~3 (bottom panel), 
the mass accretion rate has double peaks per orbit throughout the simulation.
Although the first peak of the mass accretion rate is remarkable 
for a developing phase of disc formation, 
it decrease with time after the disc is developed ($t\ga20$).
The value of the first peak could include
an artificial effect since it is related to the size of the inner
simulation boundary $r_{in}=1.0\times10^{-2}a$.


\subsection{Mode strength}

As described in Section~\ref{sec:nonaxis}, 
the one-armed spiral wave is excited by the
ram pressure due to phase-dependent mass-transfer 
from the Be disc regardless of the simulation parameters.
The strength of the mode can be defined in various ways by decomposing
the surface density distribution into Fourier components which vary as
$\exp{(im\phi)}$ with $m$ being the azimuthal harmonic number. In
paper~II, we defined it after \citet{lu}  by integrating each Fourier
component of the surface density over the whole disc and then
calculating its amplitude [see eq.~(2) of paper~II]. If the spiral wave
is, however, tightly winding, this method will give significantly
underestimated amplitudes. To avoid it, we define here the strength
$S_m$ of the mode $m$ as follows.

\begin{equation}
S_{m}(t) = \frac{1}{M_{d}}\int_{r}dr(S_{\cos, m}^{2}(r,t) + S_{\sin, m}^{2}(r,t))^{1/2},
\label{eq:mode2}
\end{equation}
where $S_{\cos, m}$ and $S_{\sin, m}$ are the azimuthal Fourier
components of the surface density given by
\begin{equation}
S_{f,m}(r,t)\equiv\frac{2}{(1+\delta_{m,0})}\int_{0}^{2\pi}rd\phi\Sigma(r,\phi,t)f(m\phi),
\label{eq:mode1}
\end{equation}
where $f$ is either $\sin$ or $\cos$ function, and
$M_{d}$ is the total disc mass given
by
\[
M_{\rm{d}}=
\int_{r}dr\int_{0}^{2\pi}rd\phi\Sigma(r,\phi,t).
\label{eq:mode3}
\]

\begin{figure*}
\resizebox{\hsize}{!}
{\includegraphics{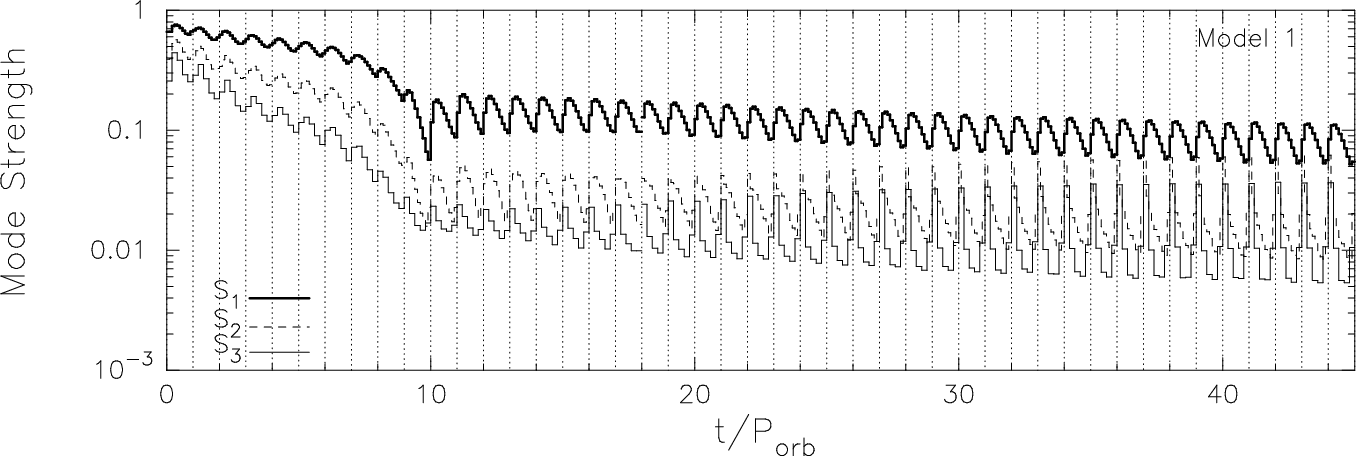}}\\
\resizebox{\hsize}{!}
{\includegraphics{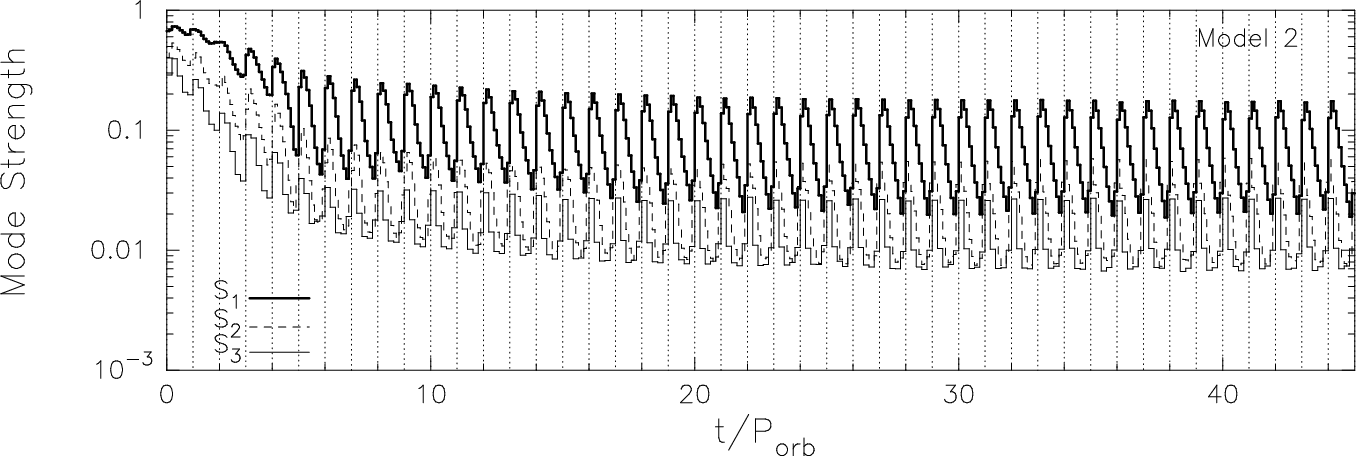}}\\
\resizebox{\hsize}{!}
{\includegraphics{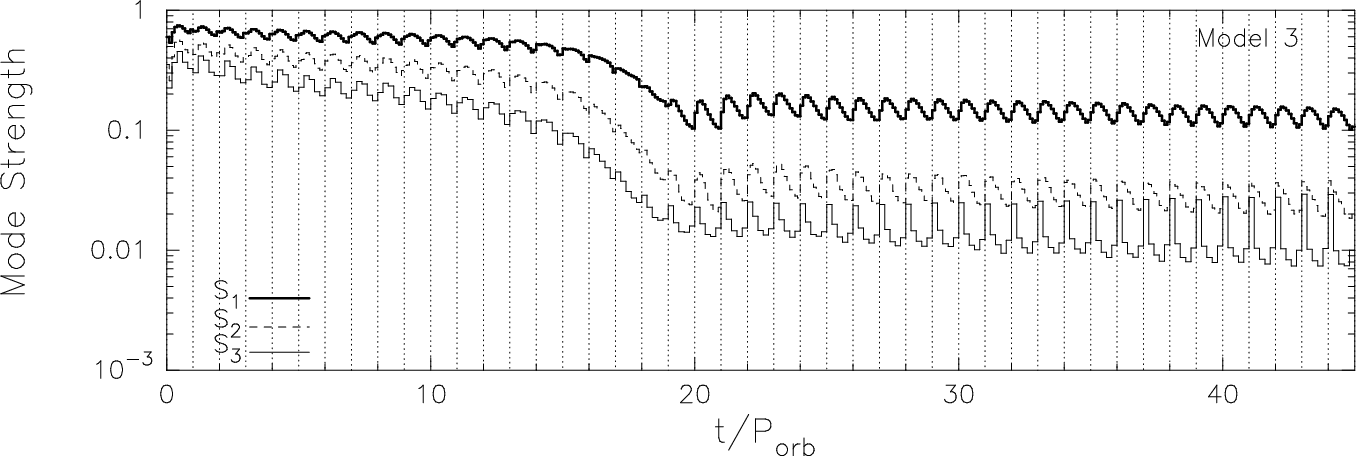}}
 \caption{
Evolution of several non-axisymmetric modes for models~1-3 with different polytropic 
exponent over the period of $0\le{t}\le45$.
In each panel, the thick solid line, the dashed line and thin solid line denote
the strengths of the $m=1$, $m=2$ and $m=3$ modes, respectively.}
 \label{fig:mode}
\end{figure*}

Fig.~\ref{fig:mode} shows the evolution of the strengths of several modes in  models~1-3.
In each panel, the solid thick line, the dotted line and the solid
thin line show the strengths of $m=1$, $m=2$ and $m=3$ modes,
respectively.
It is noted from the figure that
the one-armed mode is always dominant among modes
excited in the disc.
In the short term, the $m=1$ mode is amplified after every periastron
passage by the ram pressure due to the phase-dependent mass transfer
from the Be disc. The mode significantly decays by the next periastron
passage.
In the long-term, the amplitude of $m=1$ mode keeps decreasing until
the disc reaches the quasi-steady state.


\subsection{Disc mass, radius and viscous time-scale}
\label{sec:discgrowth}

\begin{figure*}
\resizebox{\hsize}{!}{\includegraphics*{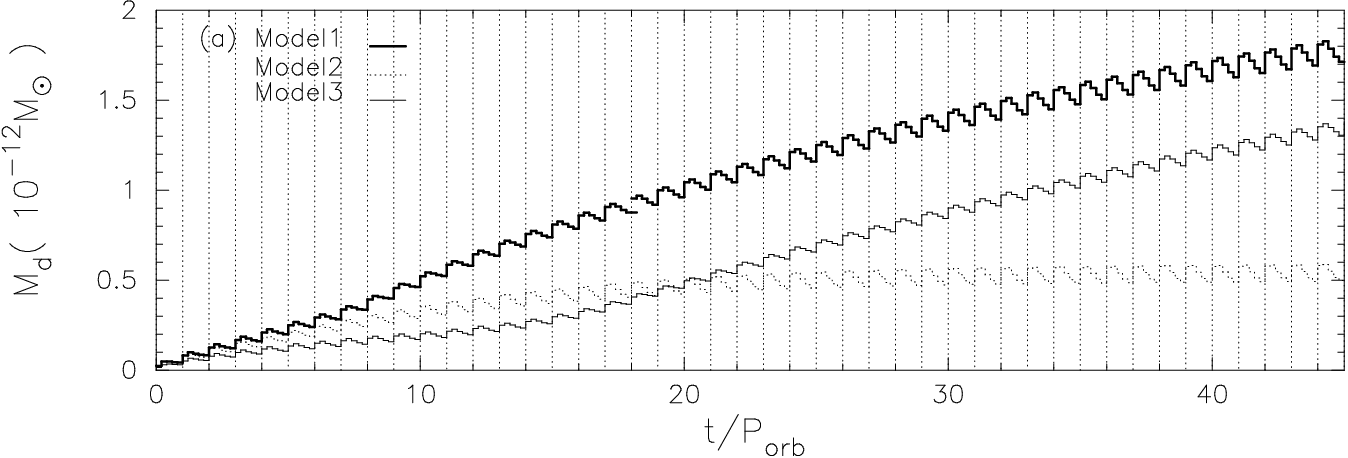}}
\resizebox{\hsize}{!}{\includegraphics*{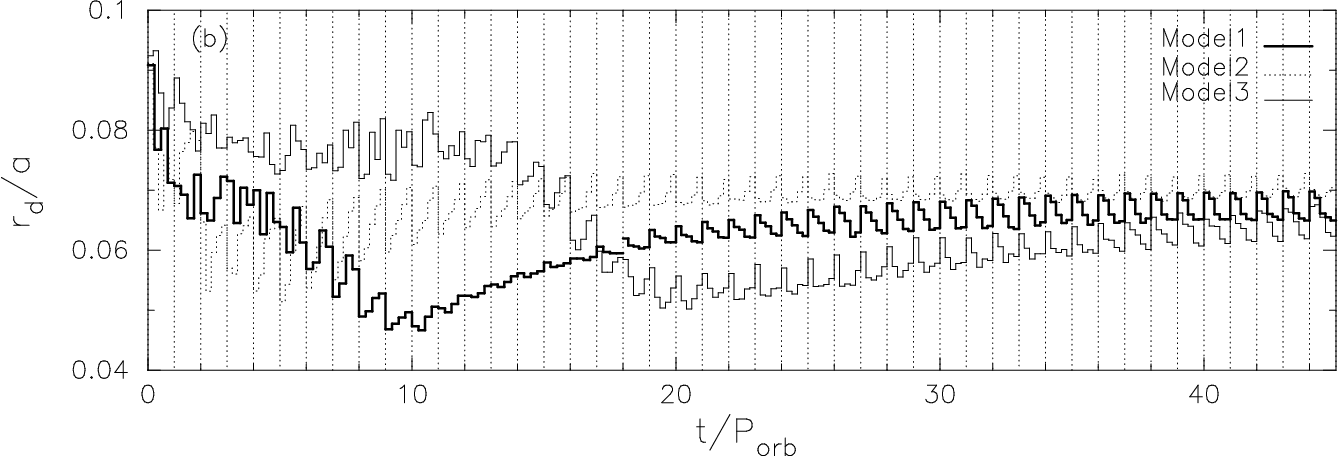}}
\resizebox{\hsize}{!}{\includegraphics*{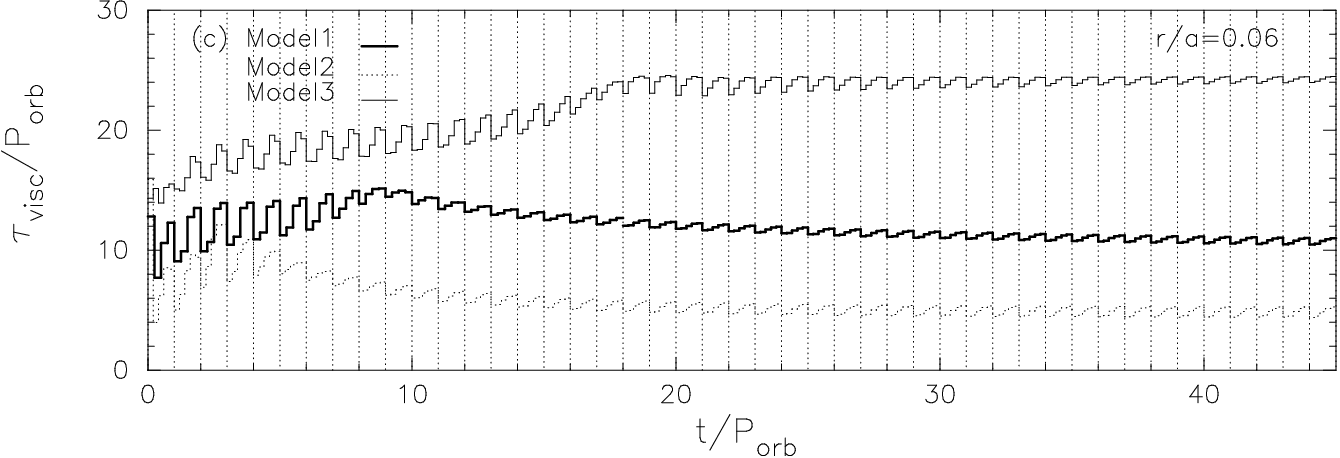}}
 \caption{
   Evolution of (a) the disc mass $M_{\rmn{d}}$ in units of $\rho_{-11}10^{-12}M_{\odot}$, 
   (b) the disc size $r_{\rmn{d}}$ normalized by the semi-major axis $a$ and
   (c) the viscous time scale of the disc $\tau_{visc}$
   in units of the orbital period $P_{\rm{orb}}$ 
    for models~1-3 with different polytropic exponents over the period of
   $0\le{t}\le45$. 
   In panel (c), 
   each viscous time-scale is estimated by using $\tau_{visc}\simeq r^{2}_{\rmn{d}}/\nu$, 
   where $\nu=\alpha{c_{s}H}$ is defined using azimuthally averaged quantities. 
         }
\label{fig:diskchara}
\end{figure*}

Fig.~\ref{fig:diskchara}(a) shows the evolution of the disc mass in
models~1-3 over the period of $0 \le t \le 45$. 
In the figure, the thick solid line, the dotted line and the thin solid line
denote the disc mass in models~1, 2 and 3,respectively.
In the short term, it modulates with the orbital phase. The disc mass starts
increasing, when the mass-supply from the Be disc begins slightly
before every periastron passage. It shows a rapid increase for a short
while, followed by a gradual decrease by accretion with no mass-supply
from the Be disc, which lasts until the next periastron passage.
In the long term, the disc
mass increases during the transition phase, 
because the averaged mass accretion rate on to the neutron star
is smaller than the averaged mass transfer rate from the Be star.
After the disc reaches a quasi-steady state, 
in which the mass accretion rate is, on average, 
balanced with the mass transfer rate, shows only 
a regular orbital modulation. For example, in model~2, the disc mass 
saturates at $\sim0.5\times10^{-12}M_{\odot}$.

In Fig.~\ref{fig:diskchara}(b), we show the evolution of the disc radius
over $0 \le t \le 45$ for models~1-3. 
As in paper~I, the disc radius is measured by applying the following
simple fitting function
\begin{equation}
    \Sigma \propto \frac{(r/r_{\rmn{d}})^{-p}}{1+(r/r_{\rmn{d}})^{q}}
    \label{eq:fitting}
\end{equation}
to the radial distribution of the azimuthally-averaged surface density
$\Sigma$. Here $p$ and $q$ are constants and $r_{\rmn{d}}$ is the
radius of accretion disc.
In the early stage of disc formation, the disc is
highly eccentric, for which the radius
$r_{\rmn{d}}$ calculated by equation~(\ref{eq:fitting}) is larger than
that for a circular disc with the same disc mass. 
Since the disc eccentricity decreases with time during the developing
phase, because of the viscous relaxation, so does the disc radius.
After the disc is circularized and enters the transition phase, the
disc radius increases with time until the disc reaches the
quasi-steady state.

Fig.~\ref{fig:diskchara}(c) shows the evolution of the viscous time-scale of the disc
over $0 \le t \le 45$ for models~1-3.
The viscous time-scale $\tau_{\rm{visc}}$ is 
estimated at the radius $r/a=0.06$ in all models using azimuthally averaged quantities.
Note that the smaller the polytropic exponent, the longer the
viscous time-scale. Thus, the viscous time-scale in model 3
($\Gamma=1$) is the longest among three models studied in this paper.
For example, at $t=45$, $\tau_{visc}\simeq11.0$, $5.5$ and $24.0P_{\rm{orb}}$ 
in models~1, 2 and 3, respectively. The disc evolution changes its trend dramatically before and after the viscous
time-scale is reached, as shown in Figs.~\ref{fig:mdots} and~\ref{fig:mode} and in Panels (b) and (c) 
of Fig.~\ref{fig:diskchara}.


\begin{figure}
\resizebox{\hsize}{!}{\includegraphics*{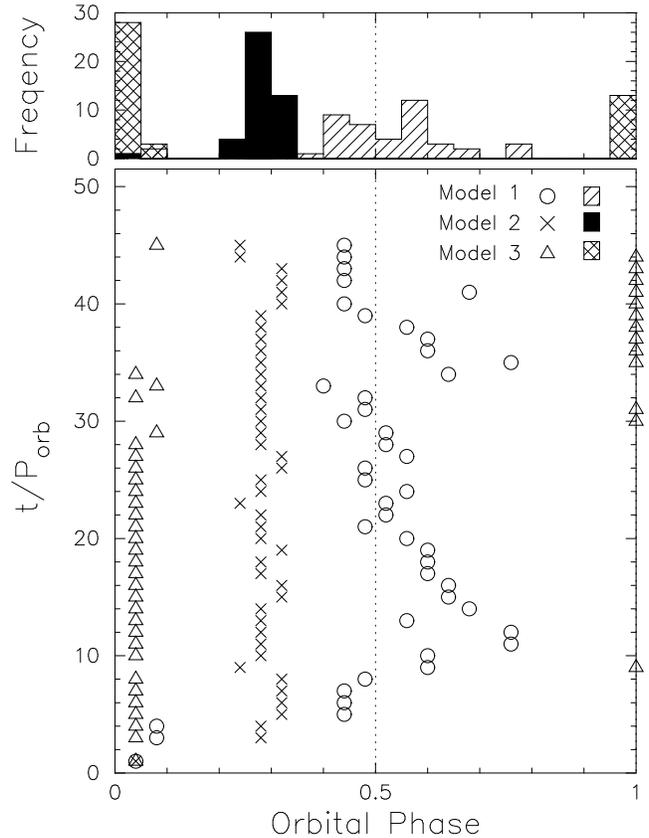}}
 \caption{
Orbital-phase dependence of 
the peak of mass-accretion rate (the lower panel)
and its frequency distribution (the upper panel).
In the lower panel, the circles, the crosses and
the triangles denote the orbital-phase dependence
 of the X-ray maxima for models~1-3, respectively.
In the upper panel, 
the histograms filled by the hatched area, the solid area and 
the cross-hatched area show 
the frequency distributions 
corresponding to the X-ray maxima for models~1-3, respectively.
         }
 \label{fig:maxphas}
\end{figure}


\section{Discussion}
\label{sec:discuss}


\begin{table*}
 \caption{
Characteristics of three evolutionary states 
of the accretion disc in Be/X-ray binaries.
'Wave-dependent phase' in the third column denotes that
 the peak phase of the wave-induced accretion 
 depends on the excitation and propagation time-scales of the one-armed wave.
} 
 \label{tbl:maxima}
 \begin{tabular}{@{}lcccccc}
  \hline
   Evolutionary state & Accretion mechanism  & 
  \begin{tabular}{c} Orbital phase of a single or \\
                     double accretion rate peaks 
  \end{tabular} \\
  \hline
  Developing phase  & Direct/Wave-induced & Periastron/Wave-dependent phase \\
  Transition phase   & Direct (only in an earlier stage)/Wave-induced & Periastron/Wave-dependent phase \\
  Quasi-steady state & Wave-induced        & Wave-dependent phase  \\
  \hline
 \end{tabular}
\end{table*}

We have performed the three dimensional simulations of the long-term evolution
of accretion disc around the neutron star in Be/X-ray binaries,
taking into account the phase-dependent mass transfer from the Be disc.
We have adopted the mass-transfer rate from a high-resolution simulation by
\citet{oka2} for a coplanar system with a short period
($P_{\rmn{orb}}=24.3\,\rmn{d}$) and moderate eccentricity $(e=0.34)$,
which targeted 4U\,0115+63, one of the best studied Be/X-ray
binaries.

We find that the disc evolves via three distinct phases, which consists of 
the first developing phase in which the disc is initially formed and developing, 
the second transition phase in which the disc is fully developed but still grows with time, and
the final quasi-steady state in which the mass-transfer rate from the Be disc 
is balanced with the mass-accretion rate onto the neutron star.

Throughout the developing phase and in an earlier stage 
of the transition phase, the mass-accretion rate has double peaks 
per orbit. The first peak, which is due to the direct accretion of 
gas particles with low specific angular momentum, occurs soon after 
periastron passage, whereas the second peak is due to the wave induced 
accretion and occurs at a phase dependent on the time-scales of excitation 
and propagation of the $m=1$ wave. 

The first peak is dominant in the developing phase. But in the earlier stage
 of the transition phase, the predominance of the first peak decreases with time, 
and finally the second peak becomes stronger.
In a later stage of the transition phase and in the quasi-steady state,
the mass-accretion rate has a single peak due to the wave induced accretion.
These characteristics are summarized in Table~\ref{tbl:maxima}.

Fig.~\ref{fig:maxphas} shows the peak phases 
of the mass accretion rate and their frequency distributions in models~1-3.
Note that the accretion rate peaks are
distributed over a wide range of orbital phase, depending on the polytropic 
exponent as well as the evolutionary state: In model~3 ($\Gamma=1$), 
in which the direct accretion of particles of low specific angular 
momentum is the major accretion mechanism, the accretion rate peaks are 
concentrated around the periastron. In contrast, in model~1 ($\Gamma=1.2$),
in which the accretion is mainly by the $m=1$ wave, the accretion rate peaks 
are distributed around the apastron with a large scatter. In model~2 ($\Gamma=5/3$),
in which the accretion is also due to the wave-induced accretion, the peak phases are
concentrated around $0.3P_{\rm{orb}}$.
The difference in the phases of accretion rate peaks between models~1, 2 and 3 
results from the difference in the major accretion mechanism, whereas 
the peak-phase difference between models~1 and 2 
is due to the difference between time-scales 
of excitation and propagation of the one-armed wave. 

The time-scale of wave propagation is roughly 
estimated by using the dispersion relation of the one-armed oscillation 
in nearly Keplerian discs \citep{ka}.
Its frequency $\omega$ is written by $\omega\sim-\Omega_{\rm K}(k_{r}H)^{2}/2$, 
where $k_{r}$ is the radial wave number. 
Then, the time-scale for a one-armed perturbation to
 travel across the disc is written 
by $\tau_{\rm{w}}\sim{r/|d\omega/dk|}\sim\alpha_{\rm{SS}}\tau_{\rm{visc}}/2\pi$.
Thus, the time-scale of wave propagation $\tau_\rmn{w}$ is obtained as
$\sim0.18P_{\rm{orb}}$ at $t=44.23$ for model~1, 
$\sim0.08P_{\rm{orb}}$ at $t=44.12$ for model~2 and
$\sim0.38P_{\rm{orb}}$ at $t=44.42$ for model~3. 
Therefore, combing the time-scales for wave excitation and propagation, 
we expect that the X-ray maxima lag behind that of the mass-transfer rate
by $\sim 0.41 P_\rmn{orb}$, $\sim 0.2 P_\rmn{orb}$ and $\sim 0.8 P_\rmn{orb}$ 
for models 1-3, respectively.
These results are in good agreement 
with the phase of the peak due to wave induced accretion 
(see Figs.~\ref{fig:mdots} and~\ref{fig:maxphas}).

The evolutionary time-scale of the disc depends on its viscous
time-scale. The viscous time-scale of the disc increases with disc
radius $r_{\rmn{d}}$, because $\tau_{\rmn{visc}}(r_{\rmn{d}}) \sim
\alpha_{\rmn{SS}}^{-1} [H(r_{\rmn{d}})/r_{\rmn{d}}]^{-2}
\Omega_{\rmn{K}}(r_{\rmn{d}})^{-1} \propto r_{\rmn{d}}^{1/2-\beta}$
with $\beta \le 0$ being the temperature gradient in the radial
direction. Thus, the larger the accretion disc, the longer the
evolutionary time-scale. Given that the size of the accretion disc
basically depends on the specific angular momentum of material
transferred from the Be disc at periastron, which is larger for longer
orbital period and/or smaller eccentricity, we expect that the
accretion disc in Be/X-ray binaries evolves most slowly in systems
with long period and low eccentricity, whereas it evolves most rapidly
in short-period, highly eccentric systems.

On the other hand, the long-term mass supply to the neutron star in
Be/X-ray binaries is controlled by the quasi-cycle of Be-star activity
(see \citealt{ne}) for $3-5$\,yr quasi-cycles of Be-disc
formation and decay in 4U\,0115$+$63). Therefore, it is likely that
accretion discs in systems with long period and low eccentricity
show only the early evolutionary state, i.e., the developing phase and
an early stage of the transition phase, whereas those in highly
eccentric, short-period systems exhibit a variety of evolutionary
states including the quasi-steady state.
The observed X-ray behaviour can be used to
probe the evolutionary state of the accretion disc in Be/X-ray
binaries.

Circinus X-1, which is a possible HMXB, 
has shown an interesting distribution of the X-ray maxima, 
although the companion is unidentified yet.
\cite{will} have examined the distribution of the X-ray maxima in Circinus X-1 
by dividing mainly into a couple of different states according to the day numbers.
In the high state, the X-ray maxima is distributed over a wide range of orbital phase
with a peak at the orbital phase $\sim0.1P_{\rm{orb}}$, 
whereas the distribution of the X-ray maxima is much more stable in the low state.
These results give an interesting suggestion on the evolutionary 
state of the accretion disc in Circinus X-1.
From the erratic and variable X-ray behaviour in the high state,
the disc is considered to be in a later stage of the transition phase.
The systems's X-ray behaviour is thus dominated by the wave induced accretion.
In the low state, the disc is considered to be in the developing phase.
Thus, the X-ray maxima resulting from the direct accretion
are concentrated at periastron, 
as shown in Table~\ref{tbl:maxima}.

\section{conclusions}
\label{sec:conclusions}

In the framework of interaction between the Be star with the circumsteller disc 
and the neutron star, 
we have carried out the three dimensional SPH simulations of the long-term evolution
of accretion disc around the neutron star 
in a coplanar Be/X binary with a short period ($P_\rmn{orb}=24.3,{\rmn{d}}$) 
and a moderate eccentricity ($e=0.34$),
taking into account the phase-dependent mass transfer from the Be disc.
Our main conclusions are summarized as follows:

\begin{enumerate}
\renewcommand{\theenumi}{(\arabic{enumi})}
\item A persistent accretion disc is formed around the neutron star in Be/X-ray binaries 
      with a short period and a moderate eccentricity.
\item The disc has a one-armed structure of transient nature. The
      one-armed density wave is excited by the ram pressure of the
      material transferred from the Be disc at each periastron
      passage, and is damped by the next periastron passage.
\item The disc evolves via three distinct phases: the first developing
      phase, the second transition phase and the final quasi-steady state. In
      the developing phase, the disc is formed and develops
      towards a nearly Keplerian disc. Then, in the transition phase, the
      approximately Keplerian disc grows with time. Finally, in the quasi-steady
      state, the mass-accretion rate on to the neutron star is on average
      balanced with the mass-transfer rate from the Be disc, and the
      disc exhibits regular orbital modulation.
\item There are two mechanisms that causes the orbital modulation of the
      accretion rate in Be/X-ray binaries: the direct accretion and the
      wave-induced accretion.
      While the former is due to the low specific angular momentum 
      of the material transferred from the Be disc, 
      the latter results from the one-armed wave
      induced by the periodically varying mass-transfer from the Be disc.
      The accretion rate has double peaks per orbit by these mechanisms in
      an earlier stage of disc evolution. In a later stage, however, it has
      only a peak arising from the wave-induced accretion.
\end{enumerate}

Be/X-ray binaries are an ideal laboratory of systems for studying
physics of accretion.
Unlike close binaries,
most Be/X-ray binaries have eccentric
orbits, the double circumsteller discs and 
can also be highly inclined systems. 
Such systems provide
us a valuable opportunity to study the effects of the
periodically-changing tidal potential and mass-transfer rate and the
inclination angle on the structure and evolution 
of the accretion flow, which hitherto have been studied
little. In addition, the knowledge on these effects are
applicable to a wide range of astrophysical objects, e.g., 
AGNs with periodic light curve such as OJ~287 which 
is considered to have binary black holes with double circumblack-hole discs 
in the central region of nuclei.
Much more work is desirable both theoretically and
observationally in order to understand this interesting and important 
group of objects.

\section*{acknowledgements}

We are grateful to the anonymous referee for constructive comments.
KH thanks Dr James R. Murray for useful comments.
KH also thanks all staffs
of the Centre for Astrophysics \& Supercomputing at
Swinburne University of Technology
for their hospitality.
The simulations reported here were performed using the facility
of the Centre for Astrophysics \& Supercomputing at
Swinburne University of Technology, Australia.
This work has been supported by Grant-in-Aid for the 21st Century
COE Scientific Research Programme on "Topological Science and Technology"
from the Ministry of
Education, Culture, Sport, Science and Technology of Japan (MECSST) and
in part by Grant-in-Aid for Scientific Research
(16540218) of Japan Society for the Promotion of Science.

\appendix

\begin{thebibliography}{}
  \bibitem[\protect\citeauthoryear{Bate, Bonnell \& Price}{Bate et al.}{1995}]{ba}
  Bate M.R., Bonnell I.A.,Price N.M., 1995, MNRAS, 285, 33
  \bibitem[\protect\citeauthoryear{Benz}{1990}]{be1}
  Benz W., 1990, in Buchler J. R.,ed.,The Numerical Modelling of Nonlinear Stellar Pulsations. 
  Kluwer, Dordrecht, p.269
  \bibitem[\protect\citeauthoryear{Benz et al.}{1990}]{be2}
  Benz W., Bowers R.L., Cameron A.G.W., Press W.H., 1990, ApJ, 348, 647
  \bibitem[\protect\citeauthoryear{Bildsten et al.}{1997}]{bi}
  Bildsten, L., Chakrabarty, D., Chiu, J., et al. 1997, ApJS, 113, 367
  \bibitem[\protect\citeauthoryear{Clarkson et al.}{2004}]{will}
  Clarkson W.I., Charles P.A \& Onyett N., 2004, MNRAS, 348, 458
  \bibitem[\protect\citeauthoryear{Frank, King \& Raine}{Frank et al.}{2002}]{fr}
  Frank J., King A.R., Raine D.J., 2002, Accretion power in Astrophysics, 
  3rd edn. Cambridge Univ. Press, Cambridge
  \bibitem[\protect\citeauthoryear{Hayasaki \& Okazaki}{2004}]{haya}
  Hayasaki K \& Okazaki A.T., 2004, MNRAS, 350, 971 (paper~I)
  \bibitem[\protect\citeauthoryear{Hayasaki \& Okazaki}{2005}]{haya2}
  Hayasaki K \& Okazaki A.T., 2005, MNRAS Letters, 360, 15L (paepr~II)
  \bibitem[\protect\citeauthoryear{Kato}{1983}]{ka}
  Kato S., 1983, PASJ, 35, 249
  \bibitem[\protect\citeauthoryear{Negueruela et al.}{2001}]{ne}
  Negueruela I., Okazaki A.T., Fabregat J., Coe M.J., Munari U \& Tomov., 2001,
  \bibitem[\protect\citeauthoryear{Lubow}{1991}]{lu}
  S H. Lubow., 1991, ApJ, 381, 268
  \bibitem[\protect\citeauthoryear{Ogilvie}{2001}]{og}
  Ogilvie G.I., 2001, MNRAS, 325, 231
  \bibitem[\protect\citeauthoryear{Okazaki et al.}{2002}]{oka2}
  Okazaki A.T., Bate M.R., Ogilvie G.I \& Pringle J.E., 2002, MNRAS, 337, 967
\end{thebibliography}
\end{document}